# Engineering the eigenstates of coupled spin-1/2 atoms on a surface

Kai Yang[1,2,3], Yujeong Bae[1,2,3], William Paul[1], Fabian D. Natterer[1,4], Philip Willke[1,2,3], Jose L. Lado[5], Alejandro Ferrón[6], Taeyoung Choi[2,3], Joaquín Fernández-Rossier[5,7], Andreas J. Heinrich[2,3,*], and Christopher P. Lutz[1,*]

[1]IBM Almaden Research Center, San Jose, CA 95120, USA
[2]Center for Quantum Nanoscience, Institute for Basic Science (IBS), Seoul 03760, Republic of Korea
[3]Department of Physics, Ewha Womans University, Seoul 03760, Republic of Korea
[4]Institute of Physics, École Polytechnique Fédérale de Lausanne, CH-1015 Lausanne, Switzerland
[5]QuantaLab, International Iberian Nanotechnology Laboratory (INL), Avenida Mestre José Veiga, 4715-310 Braga, Portugal
[6]Instituto de Modelado e Innovación Tecnológica (CONICET-UNNE), and Facultad de Ciencias Exactas, Naturales y Agrimensura, Universidad Nacional del Nordeste, Avenida Libertad 5400, W3404AAS Corrientes, Argentina
[7]Departamento de Física Aplicada, Universidad de Alicante, San Vicente del Raspeig 03690, Spain

* Corresponding authors:  A.J.H. (heinrich.andreas@qns.science) and C.P.L. (cplutz@us.ibm.com)

**Abstract:** Quantum spin networks having engineered geometries and interactions are eagerly pursued for quantum simulation and access to emergent quantum phenomena such as spin liquids. Spin-1/2 centers are particularly desirable because they readily manifest coherent quantum fluctuations. Here we introduce a controllable spin-1/2 architecture consisting of titanium atoms on a magnesium oxide surface. We tailor the spin interactions by atomic-precision positioning using a scanning tunneling microscope (STM), and subsequently perform electron spin resonance (ESR) on individual atoms to drive transitions into and out of quantum eigenstates of the coupled-spin system. Interactions between the atoms are mapped over a range of distances extending from highly anisotropic dipole coupling, to strong exchange coupling. The local magnetic field of the magnetic STM tip serves to precisely tune the superposition states of a pair of spins. The precise control of the spin-spin interactions and ability to probe the states of the coupled-spin network by addressing individual spins will enable exploration of quantum many-body systems based on networks of spin-1/2 atoms on surfaces.

Building networks of spin-1/2 objects with adjustable interactions represents a versatile approach for quantum simulation of model Hamiltonians [1, 2] because it provides direct experimental access to quantum emergent phenomena, such as topologically generated gapped excitations [3], spin liquids [4] and anyon excitations [5]. However, the precise control of spin interactions and integration beyond a few spins, while maintaining the ability to address individual spins, remains notoriously challenging [6]. Atomically engineered spin networks on surfaces, such as coupled atomic dimers, chains [7, 8], ladders [9] and arrays [10], provide a bottom-up realization of tailored spin systems, by using STM to position and address individual atoms [9, 11]. Atoms with large spin $S$ generally exhibit strong magnetocrystalline anisotropy that results in Ising-like interactions [4, 12]. In contrast, quantum fluctuations scale in proportion to $1/S$, so they are maximal for the smallest possible spin, $S = 1/2$ [4].

Spins interact via exchange and dipolar interactions. At the scale of a few coupled spins, short-range exchange coupling can give rise to magnetic ordering such as magnetic bistability [9, 13] and quantum many-body states [7, 14]. Using STM, exchange interactions have been determined by tunneling spectroscopy [7, 15], magnetization curves [10, 16] and relaxation times [17]. The recent introduction of single-atom ESR [18] increased the energy sensitivity sufficiently to allow measurements of the relatively weak, long-range dipolar interactions between high-spin magnetic atoms on a surface [12, 19].

Here we use Ti atoms adsorbed on an MgO film to realize a versatile spin-1/2 system. By combining STM with ESR, we demonstrate the ability to engineer the eigenstates and probe the quantum states of individual and pairs of Ti spins, the building blocks for simulating quantum magnetism. We control the spin-spin interactions, as well as the local magnetic field applied on individual spins, to tune the superpositions that form the eigenstates.

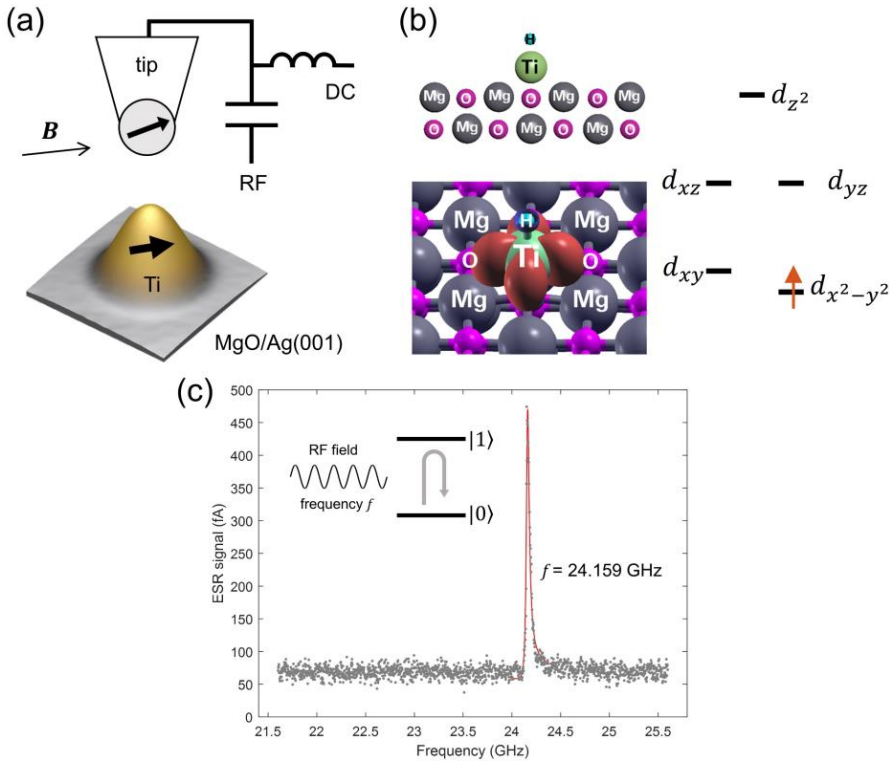

FIG. 1. ESR of a single Ti atom on MgO. (a) Schematic of the measurement set-up showing an STM image of a hydrogenated Ti atom on bilayer MgO on Ag (001), and a magnetic tip. The black arrows indicate the orientations of the magnetic moments. A radio-frequency voltage is applied to drive ESR of Ti. (b) Left panels: ball model of hydrogenated Ti on MgO, and calculated spin density. Right panel: schematic of the orbital occupancy of the $3d^1$ configuration. (c) ESR spectrum of single Ti atom. The peak is fitted to an asymmetric Lorentzian (equation (S1)) ($V$ = 50 mV, $I$ = 10 pA, $V_{RF}$ = 18 mV, $T$ = 1.2 K).

Our experiment [Fig. 1(a)] employs a bilayer MgO film grown on Ag(001) [20] in order to decouple the Ti spins from the metal substrate. The Ti atoms adsorb at oxygen-top sites, where they have $S$ = 1/2, as determined by tunneling spectroscopy [Fig. S3] and ESR measurements (below). Considering that hydrogen is the predominant component of the residual gas, and the high affinity of Ti for H in various environments [21], it is likely the Ti atom is hydrogenated. Our DFT calculations show that while clean Ti on MgO has spin $S$ = 1, hydrogenated Ti has $S$ = 1/2 [Fig. 1(b)]. The orbital moment of hydrogenated Ti is quenched, resulting in a spin-1/2 atom. Here, we focus only on the hydrogenated Ti atom species, and refer to it below simply as Ti. An external magnetic field $B$ of 0.9 T sets the Zeeman splitting of the Ti spins. The field direction is mostly in the plane of the surface, and its in-plane component is along the [110] direction of the MgO lattice [Figs. 2(b) and 2(c)]. The two spin states have spin projection −1/2 and +1/2 and are denoted |0⟩ and |1⟩ respectively.

We are able to perform single-atom ESR on individual Ti spins [Fig. 1(c)], which are driven and sensed electrically, similar to the ESR of Fe atoms [18]. Transitions between |0⟩ and |1⟩ states of the Ti spin under the tip are driven resonantly by applying a radio-frequency (RF) voltage [Fig. 1(a)]. The change of state populations is detected by tunnelling magnetoresistance when the RF frequency matches the energy difference between spin states [18]. The ESR transition is likely driven by an effective time varying magnetic field arising from the magnetic tip due to the motion of the Ti atom caused by the oscillating electric field [22]. Spin resonance of Ti at typical conditions senses spin interactions with an energy resolution of ∼0.01 μeV. It also shows a phase coherence time $T_2 \approx 100$ ns and a spin-flop time of 1.1 μs (Supplemental Section 6), comparable to those of Fe on MgO [18].

The spin Hamiltonian $H = H_{Zee} + H_{int}$ describing the spin interactions of two Ti adatoms consists of two parts (see Fig. 2(a) and Supplemental Section 7):

$$H_{Zee} = \gamma \hbar S_{1z} (B + B_{tip}) + \gamma \hbar S_{2z} B$$

$$H_{int} = J\, \mathbf{S_1} \cdot \mathbf{S_2} + D\, (3S_{1z} S_{2z} - \mathbf{S_1} \cdot \mathbf{S_2})$$

The Zeeman term $H_{Zee}$ represents the interaction of each spin with the magnetic field $B$, where $S_i = (S_{ix}, S_{iy}, S_{iz})$ is the spin operator of atom $i$, and $\gamma$ is the gyromagnetic ratio. The direction of the uniform external field $B$ is defined as $z$. The Ti spin under the tip ($\mathbf{S_1}$) experiences an additional local effective magnetic field ($B_{tip}$) due to exchange coupling to the magnetic tip [17, 22]. The interaction Hamiltonian $H_{int}$ describes both the exchange ($J$) and dipolar ($D$) couplings between the two Ti spins, which results in a correlation of their spin orientations. In the Hamiltonian $H_{int}$, we have adopted the secular approximation since the Zeeman energy is much larger than the dipolar coupling [23]. The dipolar coupling is $D = \mu_0 \mu_{Ti}^2 (1-3\cos^2\theta)/2\pi r^3$, where $\theta$ is the angle between connecting vector $\hat{r}$ and the direction of the applied magnetic field, $r$ is the Ti-Ti distance, and $\mu_{Ti}$ is the Ti magnetic moment.

The interaction strengths $J$ and $D$ can be controlled by adjusting the relative spatial positions of the two Ti spins using STM manipulation. We fabricated different Ti dimers of well-defined interatomic distances and orientations on the MgO lattice [Fig. 2(b), top panels]. ESR spectra taken with the tip positioned above one of the Ti atoms show two peaks [Fig. 2(b), bottom panels], corresponding to the two thermally occupied spin states of the coupled atom [19]. The splitting between the peaks $\Delta f = (J + 2D)/\hbar$ offers a precise measurement of the magnetic interaction strength, which strongly depends on the relative spatial positions of the atoms (Supplemental Section 7) [12, 19, 24].

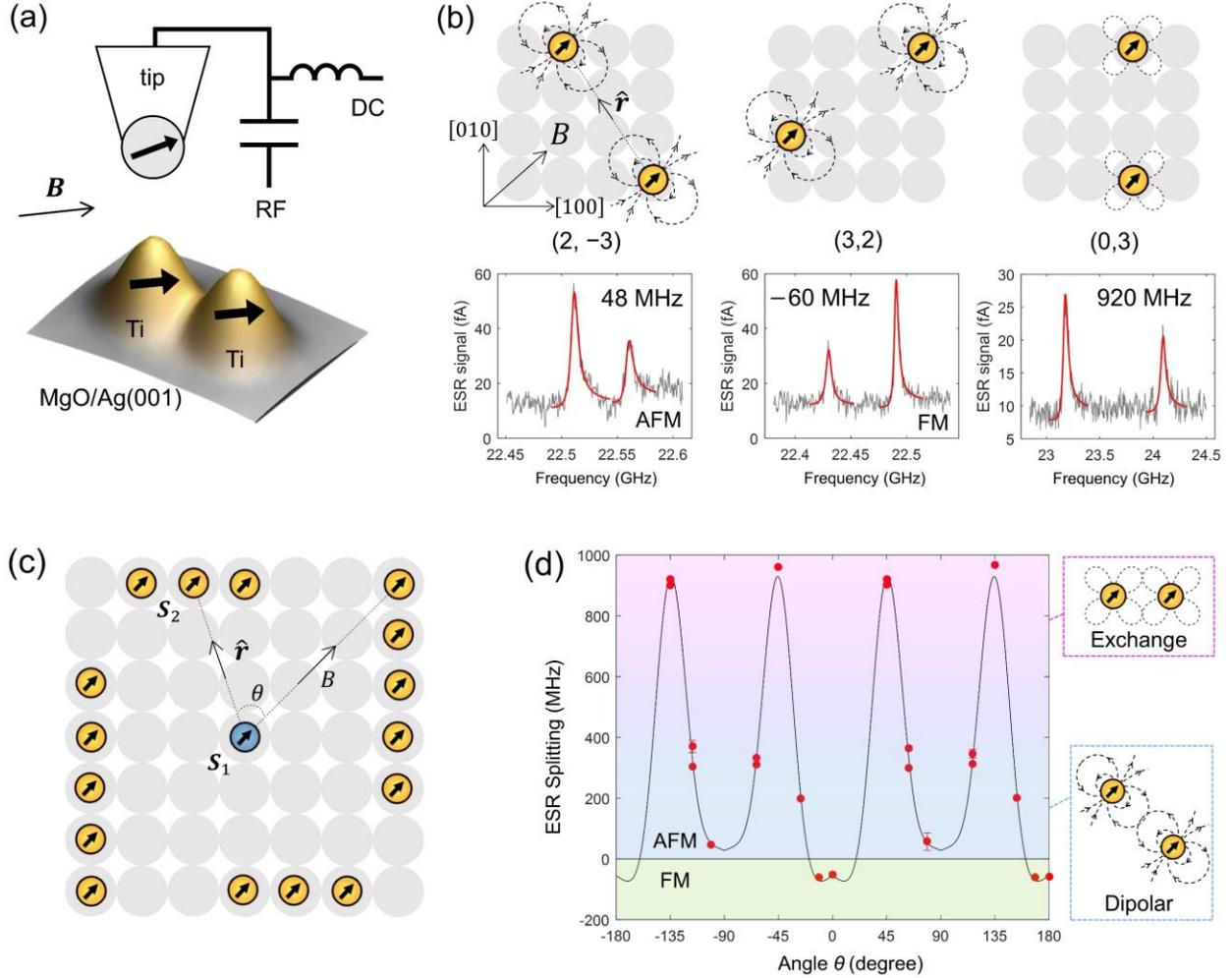

FIG. 2. Engineering magnetic couplings between two Ti atoms. (a) Schematic of the measurement of Ti-Ti couplings. (b) Top panels: positions of assembled Ti dimers are labeled ($n$, $m$) giving the number of unit cells separating them in increments of the oxygen lattice (Lattice constant: 2.88 Å). Grey circles represent oxygen atoms. Bottom panels: corresponding ESR spectra ($V$ = 50, 50, 40 mV, $I$ = 1, 1, 7 pA, $V_{RF}$ = 22, 22, 30 mV, $T$ = 1.2 K). Measured ESR splitting is shown. (c) Positions of all measured Ti dimers on MgO. Each of the yellow Ti constitutes one position relative to the center (blue) Ti. Note that only 8 dimers are needed to be assembled in order to measure the 16 unique relative positions shown in Fig. 2(c). (d) ESR splitting of 11 assembled dimers as a function of azimuthal angle $\theta$ in (c). Negative values correspond to ferromagnetic (FM) coupling. The black curve is the fit to the model Hamiltonian $H$ (see also Fig. S7(a)), tracing along a square that contains the yellow Ti in (c).

At larger atom separations, the anisotropic dipolar coupling is dominant and can be tuned from antiferromagnetic (AFM) to ferromagnetic (FM) by positioning Ti atoms at different orientations. As shown below, the eigenstates here are well described as Zeeman product states, since the Zeeman energy difference of two Ti spins due to $B_{tip}$ is much larger than the dipolar coupling. In the dimer (2, −3) in Fig. 2(b), the spins are almost perpendicular to the connecting spatial vector $\hat{r}$. This yields a positive $D$, favoring AFM coupling. Consequently, less energy is required to flip the spin under the tip when the coupled atom is in its ground state, and the taller ESR peak appears at lower frequency. The dimer (3, 2) in Fig. 2(b) has identical interatomic distance (10.4 Å) but different orientation. In this case, we find a FM interaction ($D$ is negative) and the taller peak accordingly appears at higher frequency. When the atoms are close enough, the Heisenberg exchange interaction starts to dominate, giving rise to a rapid increase in the coupling strength as the separation is reduced (dimer (0, 3) in Fig. 2(b)). The taller peak is seen at lower frequency, which indicates that the exchange coupling is AFM ($J$ is positive). Note that since we only observe two peaks in the ESR spectra of a dimer in the dipolar coupling regime, the Ti spin should have only two possible directions, and thus has a spin $S = 1/2$.

We obtain the coupling parameters $J$ and $D$ from a fit of the ESR splitting of all 11 dimers [Fig. 2(d)]. Fitting results [Figs. 2(d) and S7(a)] using isotropic exchange coupling $J = J_0 \cdot \exp(-(r-r_0)/d_{ex})$ yield a decay constant of $d_{ex} = 0.40 \pm 0.02$ Å, and a coupling strength of $0.97 \pm 0.03$ GHz at $r = r_0 = 8.64$ Å (three lattice constants). This decay constant implies a reduction of the exchange interaction by a factor of ~12 when the distance is increased by 1 Å. This behavior is comparable to the characteristic decay of exchange interaction across a vacuum gap (see Fig. 3(a)) [17, 25], which suggests that the Ti atoms are coupled through vacuum rather than by coupling mediated by the substrate conductor.

The fitting [Fig. 2(d)] also yields the moment $\mu_{Ti} = \gamma\hbar/2 = 0.99 \pm 0.11\ \mu_B$ [19], consistent with the moment of a free electron. A two-dimensional map of the fitted coupling strength [Fig. S7(b)] exhibits the expected mirror symmetry with respect to $B$ and reveals both FM and AFM regions.

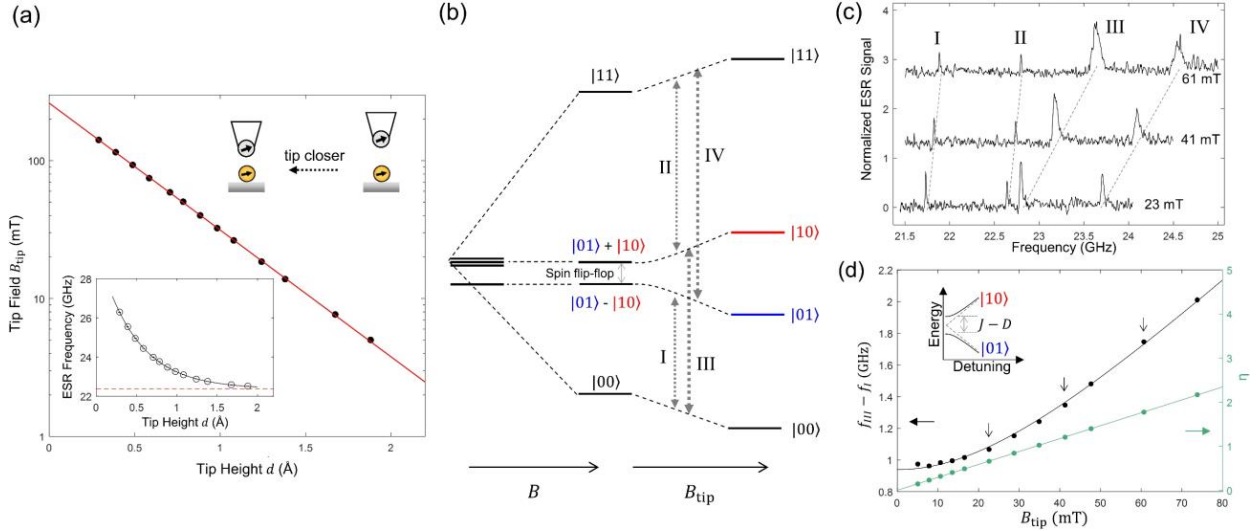

FIG. 3. Tuning the quantum eigenstates of two coupled spins. (a) Effective tip magnetic field as a function of tip height. Zero tip height corresponds to the junction resistance of 1 GΩ ($V$ = 40 mV, $I$ = 40 pA, $T$ = 1.2 K). Inset: ESR frequency of an isolated Ti atom as a function of tip height. The asymptotic value (22.37 GHz) describing the absence of the STM tip is indicated by the red dashed line. The solid lines are exponential fits. (b) Schematic energy level diagram of the two Ti spins as a function of $B$ and $B_{tip}$ for given $J$ and $D$. (c) ESR spectra on one of the Ti spins of the (0, 3) dimer at three different tip fields ($V$ = 40 mV, $I$ = 10, 7 and 4 pA, $V_{RF}$ = 30−40 mV, $T$ = 1.2 K). Spectra are normalized with respect to peak III and vertically offset for clarity. (d) $f_{III} - f_I$ as a function of $B_{tip}$ (black points). The black curve is a fit to the Hamiltonian $H$ in the main text. Arrows show positions of the spectra in (c). The green curve is the calculated interaction ratio $\eta$. Inset shows a schematic of the avoided level crossing.

We gain additional control of the spin Hamiltonian of the Ti pairs by using the effective magnetic field from the STM tip ($B_{tip}$) applied to any selected atom, in a manner similar to Ref. [17]. $B_{tip}$ is calibrated on an isolated Ti atom by measuring the ESR frequency as a function of the tip-atom distance [Fig. 3(a), inset]. The ESR frequency corresponds to the total Zeeman energy due to both $B_{tip}$ and $B$. We determine the interaction with the tip by subtracting the asymptotic value for infinite tip-sample distance. We find that tip-induced frequency shift has an exponential dependence, suggesting an exchange interaction between tip and the Ti. The exchange interaction has an exponential decay constant of 0.47 ± 0.01 Å (which varies by ~10% for different tips) and gives rise to an effective $B_{tip}$ ranging from 5–140 mT [Fig. 3(a)].

We now consider the influence of $B_{tip}$ on the quantum eigenstates of a spin-1/2 pair. We choose Zeeman product states, $|00\rangle$, $|01\rangle$, $|10\rangle$ and $|11\rangle$ as the basis. The interaction Hamiltonian can be rewritten as $H_{int} = (J+2D)S_{1z}S_{2z} + (J-D)(S_{1x}S_{2x}+S_{1y}S_{2y})$, where the $S_{1z}S_{2z}$ term shifts the energy levels of the four basis states, and the flip-flop term ($S_{1x}S_{2x}+S_{1y}S_{2y}$) causes the superposition of the states $|01\rangle$ and $|10\rangle$. For this Hamiltonian, states $|00\rangle$ and $|11\rangle$ are eigenstates, but in general, states $|01\rangle$ and $|10\rangle$ are not. The other two quantum eigenstates are:

$$|+\rangle = \cos(\xi/2)\,|10\rangle + \sin(\xi/2)\,|01\rangle$$

$$|-\rangle = \sin(\xi/2)\,|10\rangle - \cos(\xi/2)\,|01\rangle$$

where the mixing parameter $\xi$ is given by $\tan\xi = 1/\eta$, and $\eta$ is the ratio between the tip-induced energy detuning ($\gamma\hbar B_{tip}$) and the spin flip-flop coupling ($J-D$). When $\eta \gg 1$, as is the case in Fig. 2(b), the eigenstates are well described as Zeeman product states [Fig. 3(b)]. In contrast, when $\eta$ is small (the flip-flop coupling is comparable to or larger than the detuning), the eigenstates are linear superpositions of $|01\rangle$ and $|10\rangle$.

Importantly, we find that the STM can drive ESR transitions between such many-body states in multi-spin structures even though the tip is positioned to interact with only one of the atoms. The allowed ESR transitions are determined by the transition matrix element (Supplemental Section 7), which is nonzero as long as the spin quantum number of the spin under the tip differs by 1 ($\Delta m_s = \pm 1$). The superposition states $|+\rangle$ and $|-\rangle$ cannot be written as a product of states of the two spins, which makes additional transitions available, for example, from the ground state to the first excited state $|-\rangle$. We can thus drive transitions into the superposition states of the coupled spin system [Fig. 3(b)]. As a result, four ESR transitions are detected [Fig. 3(c)]. Note that the ESR transitions between triplet state $|00\rangle$ (or $|11\rangle$) and singlet state $|01\rangle-|10\rangle$ [Fig. 3(b)] are forbidden in traditional spin resonance, where a global time-varying magnetic field is used to drive spins [24].

The energy difference between the two superposition states can then be directly measured by the frequency difference between peaks III and I [Figs. 3(b), 3(c) and S8]. As we lower the tip-induced magnetic field, the frequency difference remains non-zero, indicating that the quantum states 'repel' each other near the avoided level crossing [Fig. 3(d), inset]. Fitting the peak splitting [Fig. 3(d)] yields a flip-flop coupling $(J-D) = 0.94 \pm 0.02$ GHz, in excellent agreement with the value of $(J-D) = 0.99 \pm 0.03$ GHz deduced for this dimer from the fitting results of Fig. 2(d). Note that the two additional ESR peaks (I and II) become more prominent compared to other peaks as $B_{tip}$ decreases [Fig. 3(c)]. This observation reflects the increased quantum superposition present in states |+⟩ and |−⟩.

We can quantify the quantum superposition present in the two eigenstates |+⟩ and |−⟩ at different $B_{tip}$ by calculating the interaction ratio $\eta = \gamma\hbar B_{tip}/|J-D|$. The precise control over the local magnetic field [Fig. 3(a)] enables us to tune the superposition in each eigenstate by controlling the ratio $\eta$ [Fig. 3(d)]: decreasing $B_{tip}$ reduces $\eta$, which results in increased state superposition. Note that for any given dimer, the frequency difference between peaks III and IV is constant, given by the fixed Ti-Ti coupling, and it is independent of the interaction ratio $\eta$ [Figs. 3(c) and S8].

With the ability to engineer the eigenstates of coupled spin-1/2 atoms and to probe the states at an energy scale of 0.01 μeV, it is now possible to study spin chains and networks that display phenomena such as topological states and fractional excitations [14, 26, 27]. The precise atom manipulation presented here provides scalability in constructing engineered spin networks.

We thank Bruce Melior for expert technical assistance and T. Greber for discussions. We gratefully acknowledge financial support from the Office of Naval Research. Y.B., P.W., T.C. and A.J.H acknowledge support from IBS-R027-D1. W.P. thanks the Natural Sciences and Engineering Research Council of Canada for fellowship support. F.D.N. appreciates support from the Swiss National Science Foundation under project number PZ00P2_167965. A.F. acknowledges CONICET (PIP11220150100327) and FONCyT (PICT-2012-2866). J.F-R. and J.L.L. thank Marie-Curie-ITN 607904 SPINOGRAPH and FCT, under the project "PTDC/FIS-NAN/4662/2014".

**References**

[1] K. Kim, M. S. Chang, S. Korenblit, R. Islam, E. E. Edwards, J. K. Freericks, G. D. Lin, L. M. Duan, and C. Monroe, Nature **465**, 590 (2010).
[2] A. Friedenauer, H. Schmitz, J. T. Glueckert, D. Porras, and T. Schaetz, Nat. Phys. **4**, 757 (2008).
[3] F. D. M. Haldane, Phys. Rev. Lett. **50**, 1153 (1983).
[4] L. Balents, Nature **464**, 199 (2010).
[5] A. Kitaev, Ann. Phys-new. York. **321**, 2 (2006).
[6] I. M. Georgescu, S. Ashhab, and F. Nori, Rev. Mod. Phys. **86**, 153 (2014).
[7] C. F. Hirjibehedin, C. P. Lutz, and A. J. Heinrich, Science **312**, 1021 (2006).
[8] A. Spinelli, B. Bryant, F. Delgado, J. Fernández-Rossier, and A. F. Otte, Nat. Mater. **13**, 782 (2014).
[9] S. Loth, S. Baumann, C. P. Lutz, D. M. Eigler, and A. J. Heinrich, Science **335**, 196 (2012).
[10] A. A. Khajetoorians, J. Wiebe, B. Chilian, S. Lounis, S. Blugel, and R. Wiesendanger, Nat. Phys. **8**, 497 (2012).
[11] A. A. Khajetoorians, J. Wiebe, B. Chilian, and R. Wiesendanger, Science **332**, 1062 (2011).
[12] F. D. Natterer, K. Yang, W. Paul, P. Willke, T. Choi, T. Greber, A. J. Heinrich, and C. P. Lutz, Nature **543**, 226 (2017).
[13] S. Krause, L. Berbil-Bautista, G. Herzog, M. Bode, and R. Wiesendanger, Science **317**, 1537 (2007).
[14] R. Toskovic, R. van den Berg, A. Spinelli, I. S. Eliens, B. van den Toorn, B. Bryant, J. S. Caux, and A. F. Otte, Nat. Phys. **12**, 656 (2016).
[15] J. Bork, Y.-h. Zhang, L. Diekhoner, L. Borda, P. Simon, J. Kroha, P. Wahl, and K. Kern, Nat. Phys. **7**, 901 (2011).
[16] F. Meier, L. Zhou, J. Wiebe, and R. Wiesendanger, Science **320**, 82 (2008).
[17] S. Yan, D.-J. Choi, A. J. BurgessJacob, S. Rolf-Pissarczyk, and S. Loth, Nat. Nanotechnol. **10**, 40 (2015).
[18] S. Baumann, W. Paul, T. Choi, C. P. Lutz, A. Ardavan, and A. J. Heinrich, Science **350**, 417 (2015).

[19]	T. Choi, W. Paul, S. Rolf-Pissarczyk, A. J. Macdonald, F. D. Natterer, K. Yang, P. Willke, C. P. Lutz, and A. J. Heinrich, Nat. Nanotechnol. **12**, 420 (2017).
[20]	W. Paul, K. Yang, S. Baumann, N. Romming, T. Choi, C. P. Lutz, and A. J. Heinrich, Nat. Phys. **13**, 403 (2017).
[21]	F. D. Natterer, F. Patthey, and H. Brune, Surf. Sci. **615**, 80 (2013).
[22]	J. L. Lado, A. Ferrón, and J. Fernández-Rossier, Phys. Rev. B **96**, 205420 (2017).
[23]	C. P. Slichter, *Principles of Magnetic Resonance* (Springer, 1996).
[24]	A. Abragam, and B. Bleaney, *Electron Paramagnetic Resonance of Transition Ions* (Oxford Univ. Press, Oxford, 2012).
[25]	R. Schmidt, C. Lazo, U. Kaiser, A. Schwarz, S. Heinze, and R. Wiesendanger, Phys. Rev. Lett. **106**, 257202 (2011).
[26]	R. Drost, T. Ojanen, A. Harju, and P. Liljeroth, Nat. Phys., DOI:10.1038/nphys4080 (2017).
[27]	B. Dalla Piazza, M. Mourigal, N. B. Christensen, G. J. Nilsen, P. Tregenna-Piggott, T. G. Perring, M. Enderle, D. F. McMorrow, D. A. Ivanov, and H. M. Ronnow, Nat. Phys. **11**, 62 (2015).